\documentstyle[twocolumn,aps,prl]{revtex}
\input epsf.sty
\begin{document}
\bibliographystyle{prsty}
\draft
\title{Spin-polarization-induced structural selectivity in Pd$_3X$
and Pt$_3X$ ($X=3d$) compounds}
\author{Z. W. Lu,$^{a}$ Barry M. Klein,$^{a}$ and Alex Zunger$^{b}$}
\address{$^{a}$Department of Physics, University of California,
Davis, California 95616}
\address{$^{b}$National Renewable Energy Laboratory, Golden, Colorado 80401}
\address{\rm (Submitted to Physical Review Letters on 20 December 1994)}
\address{\mbox{ }}
\address{\parbox{14cm}{\rm \mbox{ }\mbox{ }
Spin-polarization is known to lead to important {\it magnetic} and
{\it optical} effects in open-shell atoms and elemental solids,
but has rarely been implicated in controlling {\it structural} selectivity
in compounds and alloys.  Here we show that spin-polarized
electronic structure calculations are crucial for predicting the correct $T=0$
crystal structures for Pd$_3X$ and Pt$_3X$ compounds.  Spin-polarization
leads to (i) stabilization of the $L1_2$ structure over the $DO_{22}$
structure in Pt$_3$Cr, Pd$_3$Cr, and Pd$_3$Mn, (ii) to the stabilization of
the $DO_{22}$ structure over the $L1_2$ structure in Pd$_3$Co and to (iii)
ordering (rather than phase-separation) in Pt$_3$Co and Pd$_3$Cr.
The results are analyzed in terms of first-principles local spin density
calculations.
}}
\address{\mbox{ }}
\address{\parbox{14cm}{\rm PACS numbers: 61.66.Dk, 71.20Cf, and 75.50.Cc}}
\maketitle

\makeatletter
\global\@specialpagefalse
\def\@oddhead{REV\TeX{} 3.0\hfill Lu et al. Preprint, 1995}
\let\@evenhead\@oddhead
\makeatother

\narrowtext

\begin{figure}
\hskip 1.5cm
\epsfysize=5.8cm
\epsfbox{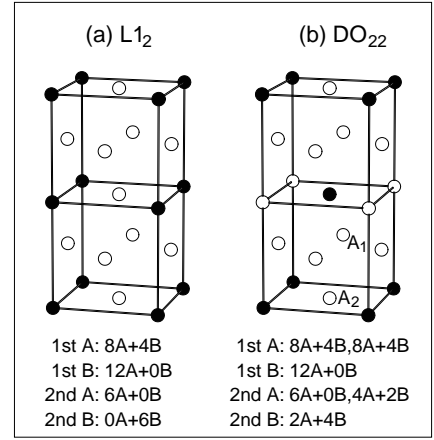}
\vskip 0.5cm
\caption{The crystal structures of the (a) $L1_2$ and (b) $DO_{22}$
structures.  The insert shows the atomic coordination about $A$
and $B$ sites in the first (1st) and second (2nd) atomic shells.
$A_1$ and $A_2$ indicate two distinct $A$ sites in the
$DO_{22}$ structure.}
\label{f-strs}
\end{figure}

Crystal structure compilations\cite{villar,mass90} reveal that the most
commonly occurring structures among intermetallic binary compounds with a 3:1
stoichiometry ($A_3B$) are the cubic $L1_2$ and the tetragonal $DO_{22}$
(Fig.~\ref{f-strs}).  The crystallographic difference between the $L1_2$
and $DO_{22}$ structures is rather subtle: the two structures have identical
{\it first} neighbor coordination (each $A$ has $8A+4B$ neighbors and each $B$
has $12A$ neighbors) while a difference exists in the {\it second} shell
(see Fig.~\ref{f-strs}).  The manner in which particular
$A_3B$ compounds select the $L1_2$ or the $DO_{22}$ configuration appears
to be rather interesting.  For example,\cite{mass90}
the $4d$ trialuminides Al$_3M$ show
the sequence $L1_2 \rightarrow L1_2 \rightarrow DO_{22}$ as
$M$ varies across the $4d$ row
Y$\rightarrow$Zr$\rightarrow$Nb, while the $3d$ palladium alloys
Pd$_3X$ show $L1_2 \rightarrow L1_2 \rightarrow DO _{22} \rightarrow L1_2
\rightarrow L1_2 \rightarrow L1_2$ as one proceeds in the $3d$ row
$X=$ Sc$\rightarrow$Ti$\rightarrow$V$\rightarrow$Cr$\rightarrow$Mn
$\rightarrow$Fe\cite{do24}
(for $X=$ Co and Ni, the systems phase-separate).
The origin of such regularities was the subject of numerous
investigations including the $d$-electron ``generalized perturbation method''
(GPM),\cite{bieb81} and first-principles
calculations.\cite{nich88,carl89,xuxx89} However,
these calculations failed to reproduce the observed structural trends.
These calculations were non-magnetic (NM, i.e., without spin-polarization).
This appeared to be a reasonable assumption, since one expects that an
alloy rich in a non-magnetic component (e.g, Pd$_3$Cr) or one without
any magnetic components (e.g, Pd$_3$V) will not have any significant magnetic
effects.  We demonstrate here that spin-polarization have a crucial influence
on the structural stability of Pd$_3X$ and Pt$_3X$ compounds: it stabilizes
the observed $L1_2$ structure over the $DO_{22}$ structure in Pt$_3$Cr,
Pd$_3$Cr, and Pd$_3$Mn, the $DO_{22}$ structure over the $L1_2$ structure in
Pd$_3$Co, and is responsible for compound formation (rather than
phase-separation) in Pt$_3$Co and Pd$_3$Cr.

The key insight to stability in compounds and alloys has traditionally
been the association of stability with low density of states (DOS)
at Fermi energy $E_F$.\cite{mott36}
\vskip 0.24cm

\begin{figure}
\hskip 0.0cm
\epsfysize=13.2cm
\epsfbox{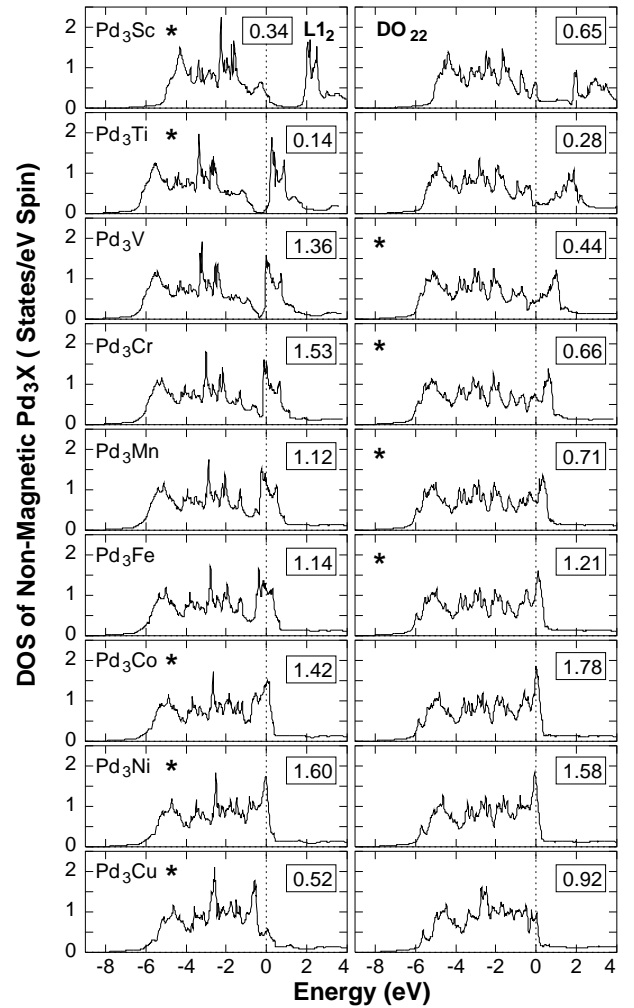}
\vskip 0.5cm
\caption{LDA calculated {\it NM} total DOS
for Pd$_3X$ in the $L1_2$ and $DO_{22}$ structures.  The inserts denote
$N(E_F)$. An asterisk denotes the more stable structure predicted
by total energy calculations in the absence of magnetic ordering.
Note that the more stable structure has a lower $N(E_F)$.
}
\label{f-dos}
\end{figure}

\begin{figure}
\hskip 0.0cm
\epsfysize=10.3cm
\epsfbox{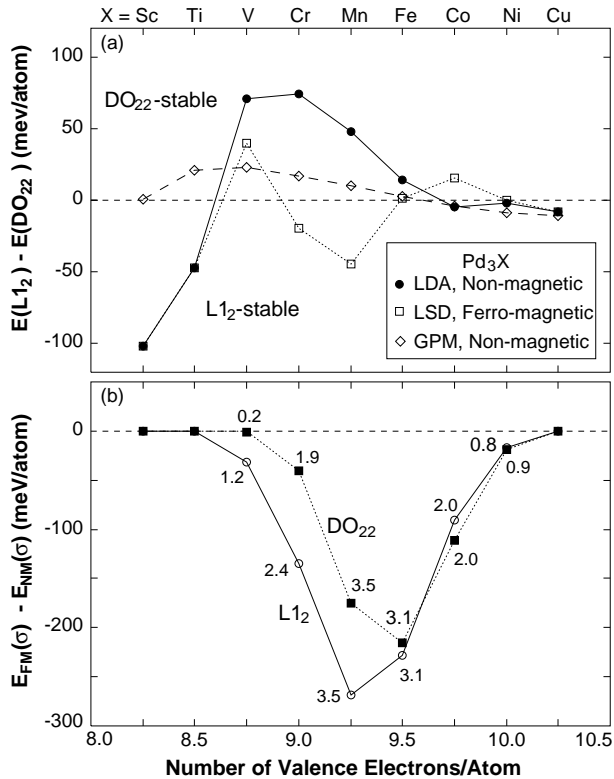}
\vskip 0.5cm
\caption{(a) LDA (present) and GPM (Ref.~\protect\onlinecite{bieb81})
calculated energy difference
$\delta E=E(L1_2)-E(DO_{22})$ [Eq.~(\protect\ref{e-reld})] for Pd$_3X$
compounds.  Note the reversal of sign for $\delta E$ due to spin-polarization
for Pd$_3$Cr, Pd$_3$Mn, and Pd$_3$Co (nearly so for Pd$_3$Fe).  Part (b) gives
the magnetization energies
$\delta M(\sigma)=E_{\rm FM}(\sigma)-E_{\rm FM}(\sigma)$
[Eq.~(\protect\ref{e-delm})] for $\sigma=L1_2$ (empty circles) and
$\sigma=DO_{22}$ (solid squares) structures and shows the local magnetic
moment for the $X$ atom (numbers above or below the symbols).
}
\label{f-energy}
\end{figure}

\noindent
Nicholson {\it et al.}\cite{nich88} noted
that for transition metal
aluminides the more stable of the two structures ($L1_2$ or $DO_{22}$)
corresponds to the one with smaller DOS at the Fermi-level, $N(E_F)$.
Local density approximation (LDA)\cite{hohe64} band structure and total energy
calculations\cite{carl89,xuxx89} have later substantiated this relation.
To examine such a relation for inter-transition
metal $A_3B$ compounds rather than aluminides
we have calculated $N(E)$ (Fig.~\ref{f-dos}) and total energy difference
$\delta E=E(L1_2)-E(DO_{22})$ [Fig.~\ref{f-energy}(a) and Table~\ref{t-tb1}]
for Pd$_3X$ with $3d$ atom $X=$ Sc through Cu
using the NM linearized augmented plane wave (LAPW)
method.\cite{sing94} We see that the structure with the lower calculated
NM total energy (Table~\ref{t-tb1}) indeed has a lower calculated NM $N(E_F)$
(Fig.~\ref{f-dos}), thus substantiating earlier trends for
aluminides.\cite{nich88,carl89,xuxx89} For example, for Pd$_3X$ with
$X=$ V, Cr, and Mn, the Fermi energy, $E_F$, falls near a DOS {\it maximum} for
$L1_2$ but near a DOS {\it minimum} for $DO_{22}$; correspondingly
$E(DO_{22})$ is lower than $E(L1_{2})$.  Unfortunately, while the magnitude
of $N(E_F)$ is indicative of the stability of the {\it calculated} structure,
these non-magnetic calculations incorrectly predict the {\it observed} stable
crystal structure
in several cases: while Pd$_3$V is correctly predicted to be
more stable in the $DO_{22}$ structure, the
{\it observed}\cite{mass90} stable structure for Pd$_3$Cr,
Pd$_3$Mn,\cite{do24} and Pd$_3$Fe is the $L1_2$ structure, not the
non-magnetically predicted $DO_{22}$ structure.  Thus, while the correlation
between the {\it calculated} quantities $N(E_F)$ versus $E(L1_2)-E(DO_{22})$
holds, it leads
to incorrect predictions for the stability of Pd$_3$Cr, Pd$_3$Mn, and Pd$_3$Fe.
The generalized perturbation method
calculations,\cite{bieb81} based on similar DOS arguments  [diamond symbols
in Fig.~\ref{f-energy})(a)] likewise predicts Pd$_3$Cr, Pd$_3$Mn, and Pd$_3$Fe
(and even Pd$_3$Sc and Pd$_3$Ti) to be stable in the $DO_{22}$ structures, in
conflict with experiment.\cite{mass90}
In this paper we explain this puzzle by noting that while a large
$N(E_F)$ indeed implies a {\it destabilizing} factor for the one-electron
(``band'') energy, it also leads (in open-shell systems) to spin-polarization
and magnetic moment formation which, in turn, is a {\it stabilizing}
factor.  Thus, despite their large $N(E_F)$ in the $L1_2$ structure (suggesting
one-electron {\it instability}), Pt$_3$Cr, Pd$_3$Cr and Pd$_3$Mn
(nearly so for Pd$_3$Fe) are correctly predicted
to be more stable in this structure once {\it spin-polarized} total energy
calculations\cite{moro93} are done.  Thus, magnetic ordering changes the
predictions of NM total energy calculations and restores agreement with
experiment.

We have calculated the total energies of Pd$_3X$, for $X=$ Sc through Cu
as well as Pt$_3$Cr and Pt$_3$Co in the $L1_2$ and $DO_{22}$ structures using
the LDA in both the spin-polarized and spin-unpolarized versions\cite{cepe80}
of the full-potential LAPW method.\cite{sing94}
In order to accurately

\begin{table}[t]
\caption{LDA calculated total energy difference
(in meV/atom) $\delta E=E(L1_2)-E(DO_{22})$ and the ferromagnetic (FM)
DOS at Fermi energy $N(E_F)$ (in states/eV spin)
for the L1$_2$ and DO$_{22}$ structures.  Note that the spin-polarization
(included in the FM state) reverses the relative stability
of non-magnetic (NM) $L1_2$ and $DO_{22}$ structures for those
compounds marked by an asterisk, thus restoring agreement with
experiment.\protect\onlinecite{mass90,do24}). PS denote
phase-separation.
}\label{t-tb1}
\begin{tabular}{lcrrdddd}
 & Expt & $\delta E_{\rm NM}$ & $\delta E_{\rm FM}$ & $N_{\rm FM}(E_F)$
 &  $N_{\rm FM}(E_F)$ \\
 & Structure &      &  $L1_2$ & $DO_{22}$ \\
\tableline
Pd$_3$Sc  & $L1_2$   & $-$102 &$-$102 & 0.34 & 0.65 \\
Pd$_3$Ti  & $L1_2$   &  $-$48 & $-$48 & 0.14 & 0.28 \\
Pd$_3$V   & $DO_{22}$&     71 &    40 & 0.64 & 0.38 \\
Pd$_3$Cr *& $L1_2$   &     74 & $-$20 & 0.57 & 0.60 \\
Pd$_3$Mn *& $L1_2$   &     48 & $-$45 & 0.59 & 0.80 \\
Pd$_3$Fe  & $L1_2$   &     14 &     1 & 0.42 & 0.34 \\
Pd$_3$Co *& PS       &   $-$5 &    15 & 0.79 & 0.46 \\
Pd$_3$Ni  & PS       &   $-$2 &     0 & 0.95 & 0.93 \\
Pd$_3$Cu  & $L1_2$   &   $-$8 &  $-$8 & 0.52 & 0.92 \\
Pt$_3$Cr *& $L1_2$   &     71 & $-$23 & 0.56 & 0.54 \\
Pt$_3$Co  & $L1_2$   &  $-$11 & $-$16 & 0.66 & 0.65 \\
\end{tabular}
\end{table}

\noindent obtain the small energy difference between two fairly
similar crystal
structures, the calculations were carried out consistently
using the same muffin-tin radii, $R_{\rm MT}$, and basis set energy cutoffs,
$K_{\rm max}$.   The Brillouin Zone summations were done using the
geometrically equivalent {\bf k}-point sampling scheme''\cite{froy}(a),
in which 20 (40) {\bf k}-points in the irreducible zone for the $L1_2$
($DO_{22}$) structure is
mapped into the same 60 special {\bf k}-points\cite{froy}(b) in the
fcc structure. We optimized the total energy as a function of volume,
as well as the $c/a$ ratio in the $DO_{22}$ structure.
The estimated LAPW error for the $E(L1_2)-E(DO_{22})$ energy difference is
$\sim 5$~meV/atom, and the neglected zero-point energy difference
between the two similar structures should be even smaller.  At zero
temperature, the {\it absolute stability of a compound in a structure
$\sigma$} with respect to phase separation is given
by its formation enthalpy $\Delta H(\sigma)$
\vskip 0.2cm

\noindent
\begin{equation}
\Delta H_\alpha(\sigma)=E_\alpha(\sigma)-x_AE_A(V_A)-x_BE_B(V_B)
\label{e-delh}
\end{equation}
where $E_A(V_A)$ and $E_B(V_B)$ are the energies of the constituents $A$
and $B$ in their respective ground states (e.g., for Pd$_3$Ni, it is
non-magnetic fcc for Pd and ferromagnetic fcc for Ni), $E_\alpha(\sigma)$
is the energy of structure $\sigma$, $\alpha=$ NM or FM denotes
whether structure $\sigma$ is
in the non-magnetic or ferromagnetic states.  The {\it relative stability}

{\it of two different ordered structures} is,
\begin{equation}
\delta E_\alpha=
E_\alpha(L1_2)-E_\alpha(DO_{22}) \;\;,
\label{e-reld}
\end{equation}
while the {\it magnetic stabilization energy of a given structure
$\sigma$ ($L1_2$ or $DO_{22}$)} is
\begin{equation}
\delta M(\sigma)=
E_{\rm FM}(\sigma)-E_{\rm NM}(\sigma) \;\; .
\label{e-delm}
\end{equation}

Table~\ref{t-tb1} and Fig. \ref{f-energy}(a) give $\delta E_{\rm NM}$ and
$\delta E_{\rm FM}$, while Fig.~\ref{f-energy}(b) shows the magnetic
stabilization energies $\delta M(L1_2)$ and $\delta M(DO_{22})$, and the
local magnetic moment $\mu_X$ on the $X$ atom calculated numerically
within the muffin-tin sphere (the value of $\mu_X$ is rather
insensitive to the small change in muffin-tin radius).

\noindent Table~\ref{t-tb2} gives the formation
enthalpies $\Delta H_\alpha(L1_2)$ and $\Delta H_\alpha(DO_{22})$ for several
systems. Figure~\ref{f-dos} depicts the NM total DOS for Pd$_3X$, where $X=$
Sc$\rightarrow$Cu, in the $L1_2$ (left panel) and $DO_{22}$ structures
(right panel).  One notices the following:

(i)  In contrast to the very similar DOS for the
$L1_2$ and $DO_{22}$ structures and the small
energy difference $E(L1_2)-E(DO_{22})$ $\sim 20$ meV/atom
predicted by the GPM,\cite{bieb81} one sees from Fig.~\ref{f-dos} a
marked difference of the DOS in the $L1_2$ and
$DO_{22}$ structures.
(a) While the DOS of the cubic $L1_2$ structure resemble that of fcc Pd,
having three major peaks, the DOS of the $DO_{22}$ structure are more
smeared, reflecting a loss of cubic symmetry in the $DO_{22}$ structure.
(b) The DOS of the more stable $L1_2$ structure for Pd$_3$Sc and Pd$_3$Ti
shows a ``pseudo-gap'' near the Fermi level absent in the $DO_{22}$ structure.
(c) In the $L1_2$ structure, the Fermi level of Pd$_3$V, Pd$_3$Cr, and
Pd$_3$Mn falls on a DOS peak (made mostly of $d$ orbitals of the $X$ atom),
while in the $DO_{22}$ structure the Fermi level falls on a relatively
flat portion of the DOS.
Indeed, these materials are more stable in the $DO_{22}$
structure in a NM LDA description.  As a result of these differences,
the values of the $N(E_F)$ (given in the inserts
of Fig.~\ref{f-dos}) and its shape near the Fermi level are strikingly
different for the $L1_2$ and $DO_{22}$ structures.

(ii) The above noted trends in the NM $N(E_F)$ induce a concomitant
magnetic stabilization energy $\delta M$ [Fig.~\ref{f-energy}(b)] :
A {\it larger} energy stabilization $\delta M$ due to
spin-polarization relates with a {\it larger} $N(E_F)$ and
with a {\it larger} localized magnetic moment $\mu_X$ on the $X$ atom
[Fig.~\ref{f-energy}(b)].
For example, while the Pd$_3X$ compounds with $X=$ Sc, Ti, and Cu
are non-magnetic (so $\delta M=0$), when $X=$ Mn and Fe, one sees
in Fig~\ref{f-energy}(b) large
energy lowering due to spin-polarization ($\delta M \sim -200$ meV/atom),
and concomitantly large magnetic moments
of 3.5 $\mu_B$ ($X=$ Mn) and 3.1 $\mu_B$ ($X=$ Fe) in both the $L1_2$ and
$DO_{22}$ structures (as a comparison, bcc Fe has a magnetic moment value
of only 2.2 $\mu_B$).  Thus, spin-polarization induces a local magnetic
moment on the ``magnetic'' $3d$-atoms with large NM $N(E_F)$ while lowering
the total energy of the compound.  In the case of Pt$_3$Cr in the $L1_2$
structure for which previous calculation exists, our calculated total
magnetic moment in the unit cell (2.6 $\mu_B$) agrees with a

\begin{table}
\caption{Non-magnetic (NM) and Ferromagnetic (FM)
formation enthalpies, $\Delta H$ [in meV/atom, Eq.~(\protect\ref{e-delh})],
of some compounds.  $\Delta H$'s are taken with respect
to the NM fcc Pd, Pt, FM fcc Ni, FM fcc Co, and anti-FM bcc Cr,
respectively.
}\label{t-tb2}
\begin{tabular}{lrrrr}
 &\multicolumn{2}{c}{$\Delta H(L1_2)$} &
  \multicolumn{2}{c}{$\Delta H(DO_{22})$} \\
 & NM & FM & NM & FM \\
\tableline
Pd$_3$Cr &    126 &   $-$9 &     51 &     11 \\
Pd$_3$Co &    155 &     64 &    160 &     49 \\
Pd$_3$Ni &     61 &     43 &     62 &     44 \\
Pt$_3$Cr &  $-$68 & $-$185 & $-$139 & $-$161 \\
Pt$_3$Co &     31 &  $-$42 &     42 &  $-$26 \\
\end{tabular}
\end{table}

\noindent previous calculation\cite{szaj92} of 2.6 $\mu_B$ and with the
experimental value\cite{burk80} of 2.5 $\mu_B$.

(iii) In the NM calculations, Pd$_3$V and Pd$_3$Cr have large DOS
peaks at $E_F$ in the $L1_2$ structure, and are concomitantly less stable
in this structure than in the low $N(E_F)$ $DO_{22}$ structure.  However,
as spin-polarization is introduced, the large DOS peak of the $L1_2$ structure
splits, so that $E_F$ now resides in a low DOS region.  This leads,
simultaneously, to the formation of larger magnetic
moments on V and Cr in the $L1_2$
structure relative to the $DO_{22}$ structure.  This selective magnetization
thus lowers the energy of the $L1_2$ structure more significantly
than in the $DO_{22}$ structure [Fig.~\ref{f-energy}(b)].

(iv) Table~\ref{t-tb1} shows that the more stable of the two structures
generally (and weakly) relates with a smaller value of the {\it ferromagnetic}
DOS at the Fermi level $N_{\rm FM}(E_F)$.  Thus, the trend of total energy
stability with $N(E_F)$ does exist, but for the spin-polarized quantities.

(v) The magnetic stabilization energy $\delta M$, reverses the relative
stability predicted by NM calculations in several cases:
spin-polarization stabilizes the experimentally observed\cite{mass90}
$L1_2$ structure of Pd$_3$Cr, Pd$_3$Mn, and Pt$_3$Cr (nearly so for
Pd$_3$Fe) over the $DO_{22}$ structure, while for Pd$_3$Co, spin-polarization
makes the $DO_{22}$ structure more stable.
The reversal of stability for Pd$_3$Co can not be observed experimentally,
since the calculated $\Delta H$ [Eq.~(\ref{e-delh}) and
Table~\ref{t-tb2}] is {\it positive}, so
Pd$_3$Co (and similarly, Pd$_3$Ni) is predicted to phase separate rather
than to order, in accord with the observed phase-separation
behaviors for Pd-Co and Pd-Ni.\cite{mass90}

(vi) Spin-polarization can stabilize ordering over
phase-separation:\cite{luxx91}
We find that in a NM description Pt$_3$Co has $\Delta H_{\rm NM}>0$
so it is predicted to phase separate, but that a strong spin-polarization
effect stabilizes the ordered $L1_2$ structure, leading to
$\Delta H_{\rm FM} <0$ in agreement with the observed ordering
behavior.\cite{mass90}
Similarly, spin-polarization stabilizes the experimentally
observed\cite{huang88} $L1_2$ structure of Pd$_3$Cr (its magnetic behavior
has, however, not been experimentally examined).  Hence, spin-polarization
not only reverses the stability of the $L1_2$ and $DO_{22}$ structures for
many compounds, but it also stabilizes an ordered ($L1_2$) structure over
phase-separation for Pd$_3$Cr and Pt$_3$Co.

(vii) Interestingly, the $\Delta H$ for Pt$_3$Cr
are negative in both the NM and FM cases, but the spin-polarization effect
gives a larger stability to the $L1_2$ structure (observed
experimentally).\cite{mass90} One also notices that the
$\Delta H$ are lower in the Pt alloys than in the corresponding Pd alloys.
The increased stability in Pt alloys has been addressed previously in Ref.
\onlinecite{luxx91}.

In summary, we have shown that a theoretically unstable non-magnetic
structures which involve magnetic atoms and possesses {\it large} $N(E_F)$,
may be stabilized by splitting the near $E_F$ peak in the DOS and forming a
local magnetic moment with a concomitant {\it lowering} of
$N(E_F)$ and the total energy.  Therefore, theoretical studies of the
stability of compounds with a large value of a NM $N(E_F)$ should be carefully
tested for magnetic ordering which can often change the predictions of the
ground state crystal structure.

We thank Professor Alan Ardell for pointing out the fact that Pd$_3$Cr
orders in the $L1_2$ structure.  ZWL and BMK acknowledge
the support by the University Research Funds of the University of California
at Davis and San Diego Supercomputer Center for comuter time.
AZ acknowledges the DOE, the office of energy research, basic
energy sciences, division of materials science for support under the contract
No. DE-AC36-83CH10093.



\begin{references}

\bibitem{villar} P. Villar and L. D. Calvert, {\it Pearson's Handbook of
Crystallographic Data for Intermetallic Phases},
(ASM, Materials Park, Ohio, 1991).

\bibitem{mass90} {\it Binary Alloy Phase diagrams}, edited by T. B. Massalski
{\it et al.} (ASM, Materials Park, Ohio, 1990).

\bibitem{do24} Pd$_3$Ti crystallizes in a hexagonal
$DO_{24}$ structure as well as in an off-stoichiometric $L1_2$ structure.
Pd$_3$Mn is observed in the $DO_{23}$ structure, which could be characterized
as a combination of
the $L1_2$ and $DO_{22}$ structures.

\bibitem{bieb81} (a) A. Bieber {\it et al.} Solid State Commun. {\bf 39}, 149
(1981);(b) {\it ibid} {\bf 45}, 585 (1983). (c) A. bieber and F. Gautier,
Acta Metall. {\bf 34}, 2291 (1986).

\bibitem{nich88} D. M. Nicholson {\it et al.}, Mat. Res. Soc. Proc. {\bf 133},
17 (1989);{\bf 186} 229 (1991).

\bibitem{carl89} A. E. Carlson and P. J. Meschter, J. Mat. Res. {\bf 4},
1060 (1989).

\bibitem{xuxx89} Xu and A.~J. Freeman, Phys. Rev. B {\bf 40}, 11927 (1989);
J. Mat. Res. {\bf 6}, 1188 (1991).

\bibitem{mott36} N. F. Mott and H. Jones, {\it The Theory of the Properties
of Metals and Alloys}, (Clarenden Press, Oxford, 1936).

\bibitem{hohe64} W. Kohn and L. J. Sham, Phys. Rev. {\bf 140}, A1133 (1965).

\bibitem{moro93} E. G. Moroni and T. Jarlborg, Phys. Rev. B {\bf 47}, 3255
(1993).

\bibitem{sing94} D.~J. Singh, {\it Planewaves,
Pseudopotentials, and the LAPW Method}, (Kluwer, Boston,
1994).

\bibitem{cepe80} We use the exchange-correlation potential of
D. M. Ceperley and B. J. Alder, Phys. Rev. Lett {\bf 45}, 566 (1980)
as parameterized by  J. P. Perdew and A. Zunger, Phys. Rev. B {\bf 23},
5048 (1981).

\bibitem{froy} (a) S. Froyen, Phys. Rev. B {\bf 39}, 3168 (1989);
(b) H. J. Monkhorst and J. D. Pack, {\it ibid} {\bf 13},
5188 (1976).

\bibitem{szaj92} A. Szajek, Acta Phys. Polo. A {\bf 82}, 967 (1992).

\bibitem{burk80} S. K. Burke {\it et al.}, J. Magn. and Magn. Mater.
{\bf 15-18}, 505 (1980).

\bibitem{luxx91} Z. W. Lu, S.-H. Wei, and A. Zunger, Phys. Rev. Lett.
{\bf 66}, 1753 (1991),  these authors have also shown that relativistic
effects can stabilize ordering over phase separation in NiPt.

\bibitem{huang88} J.~C. Huang, A.~J. Ardell and O. Ajaja,
J. of Mater. Sci. {\bf 23}, 1206 (1988).


\end{references}
\end{document}